\documentclass[twoside]{article}
\usepackage{fleqn,espcrc2}

\usepackage{epsfig}

\newcommand{\AmS}{{\protect\the\textfont2
  A\kern-.1667em\lower.5ex\hbox{M}\kern-.125emS}}

\title{Neutrino oscillations in the early Universe}

\author{D. Kirilova\address{Institute of Astronomy, 
        Bulgarian Academy of Sciences, 
        Tsarigradsko Shosse 72, Sofia, Bulgaria}
        and 
        M. Chizhov\address{Centre for Space Research and Technologies,
        University of Sofia, 1164 Sofia, Bulgaria}%
 \thanks{CERN, CH-1211 Geneva 23, Switzerland.}}
       
\begin{document}

\begin{abstract}
   We discuss  the oscillations effects on neutrinos in the early
Universe  
and update  the cosmological constraints on the oscillation parameters. 
It is shown that sterile LOW solution to the solar neutrino problem is
almost completely excluded from cosmological
nucleosynthesis considerations.
Two possibilities for the relaxation of this constraint are discussed:
high primordial $^4\!He$ 
yield  and a relic lepton asymmetry
present at nucleosynthesis epoch. The numerical analysis proved that     
$Y_p=0.25$  only relaxes the constraint on LOW solution, while $L\ge
10^{-5}$ is capable to remove it.

\end{abstract}

\maketitle

\section{Neutrino oscillation effects}

Neutrino oscillations may play a considerable role in the early Universe.
Cosmological nucleosynthesis (CN), baryogenesis, microwave background
radiation, large scale structure formation, dark matter, lepton asymmetry,  
etc. could be essentially influenced by nonstandard neutrino properties,
like nonzero mass and oscillations, the presence of additional neutrino
types,  etc.

 The oscillations effect depends on the type of oscillations:
oscillation channels, resonant transitions, the degree of equilibrium of 
 oscillating neutrinos. Neutrino oscillations
are capable to

(i) bring additional light particles into 
equilibrium~\cite{do},

(ii) deplete the neutrino  number 
density~\cite{bd1},

(iii) distort the neutrino energy spectrum~\cite{dpk,NU96}
and

(iv) affect neutrino-antineutrino asymmetry~\cite{NU96,create}. 

\noindent The effects (i) and (iv) are typical for active-sterile
oscillations, and
(ii)--(iv) are most considerable for nonequilibrium active-sterile 
oscillations.  All these
play crucial role for neutrino involved processes in the early Universe.

\section{Cosmological constraints on neutrino oscillation parameters}
 
Special attention is due to cosmological nucleosynthesis, which provides 
the strongest constraints on neutrino oscillation
parameters~\cite{ekt,do00,PR,res,dubna}. 
Primordial yield of $^4\!He$ can
be calculated with great accuracy within the standard CN~\cite{helium}.     
Helium-4 values, extracted from observation, although perhaps suffering
from great systematic errors (of the order  0.05), are still the most
reliable among the relic light element yields. Hence, the most
reliable cosmological constraints are obtained in studies of that element.  

First CN constraints, based on effect (i) were provided in
refs.~\cite{do,bd1}. The best constraints available now on 
$\nu_{\mu,\tau}\leftrightarrow \nu_{s}$, accounting for (i) and partially
for (ii), are provided in ref.~\cite{ekt}. They should be updated for 
the effects (iii) and (iv). 
\footnote{An attempt to account for (iii) was made in
ref.~\cite{shif99}, 
however, the work contains discontinuity in the results for
the nonresonant and the resonant
cases at maximal mixing, and hence is not reliable.}

The constraints on
$\nu_e\leftrightarrow \nu_{s}$ have been recently updated. 
An analytical study of the nonresonant case, accounting for (iii)
to some acceptable approximation was provided~\cite{do00}. 
Precise numerical analysis of  
nonresonant~\cite{PR} and  resonant~\cite{res} cases were performed. 
 Exact kinetic equations for neutrino density
matrix in
momentum space were used to describe
oscillating neutrinos in the high temperature Universe. Precise
account for (i)--(iv) effects and  
selfconsistent analysis of neutrino and nucleons
evolution during the CN epoch was made~\cite{PR,res,dubna}. 

In fig.~1 the updated constraints on 
nonresonant and resonant  $\nu_e \leftrightarrow \nu_s$  for different
primordial $^4\!He$ values are plotted.
The analytical fits to the exact constraints for primordial
$^4\!He$ $Y_p=0.24$ are:
\begin{eqnarray*}
\delta m^2 (\sin^22\vartheta)^4\le 1.5 \times10^{-9} {\rm eV}^2
&&\delta m^2>0, \\
{\rm and}~~~~~|\delta m^2| < 8.2\times 10^{-10} {\rm eV}^2 
&&\delta m^2<0
\end{eqnarray*}
at large mixing angles.
These constraints
are an order of magnitude stronger at large mixings than the previous 
due to the precise account of (ii) and (iii). In the resonant
case they are less restrictive at small
mixings, due to the account of (iv) -  
neutrino-antineutrino asymmetry  generated in oscillations.
\begin{figure}[htb]
\epsfig{file=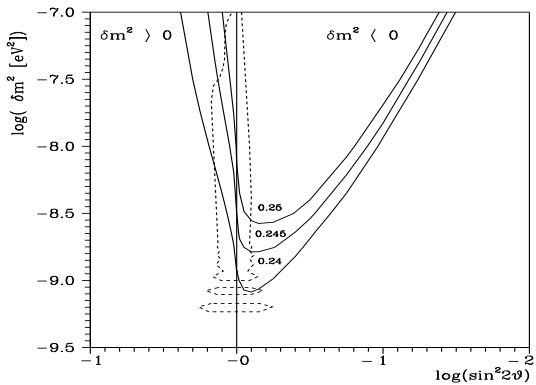,height=5cm,width=7.0cm}
\vspace{-15pt}
\caption{Cosmological constraints for $\nu_e\to\nu_s$ are presented
for $Y_p=0.24,0.245,0.25$. The dotted curve shows solar neutrino LOW
solution.}
\label{1}
\vspace{-10pt}
\end{figure}

According to these constraints, besides active-sterile
LMA solution also LOW solution
to the solar
neutrino problem is almost completely excluded. 
In fig.~1 the LOW region, plotted by a dotted curve, is taken from 
ref.~\cite{suzuki}. 

 \section{Relaxation of CN constraints}

 We have studied two possibilities for relaxing  
 the cosmological constraints:
 
{\bf 1. Higher $Y_p$.} Having in mind the large systematic 
error of $Y_p$ extracted from observations, $Y_p>0.24$ 
looks possible. Therefore, we  have calculated  iso-helium contours
$Y_p=0.245,0.25$  and compared them with the LOW solution.  
CN constraints then are relaxed, however, 
even  $Y_p=0.25$ cannot remove completely 
the constraints on LOW solution (fig.~1).\footnote{Mind that the account
only of (i) 
does not constrain oscillations for high $Y_p$. To obtain the exact
constraints the precise account of (ii)--(iv) and the selfconsistent study
of oscillations and CN is obligatory.} 

{\bf 2. Relic lepton asymmetry $L$},
 present before CN, could relax the bounds. We have studied  small
asymmetries $L<<0.01$ that do not effect directly CN kinetics. 
They influence CN {\it indirectly} via oscillations:
 
(a) effecting  neutrino number densities, 

(b) neutrino spectrum distortion and 

(c) neutrino oscillation pattern (suppressing or enhancing
oscillations). 

\noindent This reflects in underproduction or overproduction of
$^4\!He$~\cite{CAPP,NP,new} in comparison with the case without $L$.

We have analyzed the effect of relic $L$ on CN with oscillations,
providing a precise selfconsistent
study of CN and oscillating neutrinos for each set 
($\delta m^2$, $\sin^22\vartheta$, $L$). The iso-helium contours
$Y_p=0.24$ for $L=10^{-10}$ and $L=10^{-6}$ are presented in fig.~2.
\begin{figure}[htb]
\epsfig{file=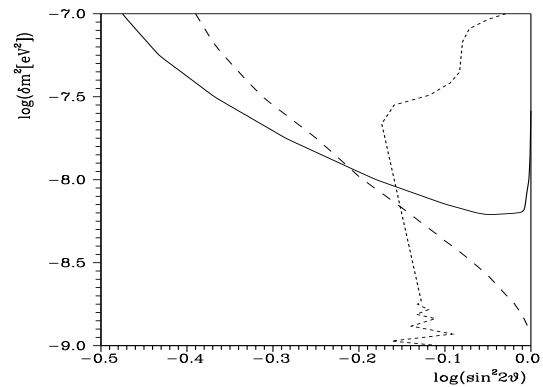,height=5cm,width=7cm}
\vspace{-15pt}
\caption{Iso-helium contours $Y_p=0.24$ for $L=10^{-6}$(solid curve) and
$L=10^{-10}$ are shown.}
\label{2}
\vspace{-10pt}
\end{figure}
Relic lepton asymmetries strengthen the bound at small $\vartheta$ and
relax them at large mixings. The 
numerical analysis for $L=10^{-6},10^{-5.5}$ showed that such 
$L$  relax the
constraints, while $L= 10^{-5}$ can remove
the CN constraints on LOW solution. 

{\bf Oscillations generated asymmetry}. Lepton asymmetry can be 
dynamically generated due to resonant 
oscillations. Oscillations generated
asymmetry can suppress oscillations and 
alleviate  CN constraints~\cite{NU96,create,new}.
We have several remarks concerning this possibility. 

Often very 
rough estimation of asymmetry growth is provided, without a precise
kinetic account of the indirect asymmetry effects on CN and without 
account for the neutrino spectrum distortion~\cite{wrong}. Moreover,
the asymmetry effect on CN is discussed  separating artificially the
processes of
asymmetry
growth and the CN:  discussing first asymmetry growth till big $L
\ge 0.01$ and afterwards exploring kinetic effect of big $L$ 
on CN.

\begin{figure}[htb]
\epsfig{file=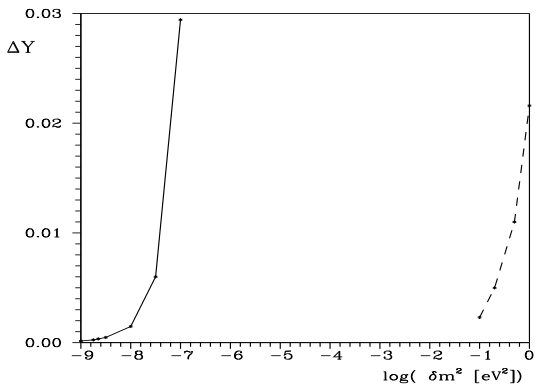,height=5cm,width=7cm}
\vspace{-15pt}
\caption{Relative increase of $Y_p$ as a function of $\delta m^2$.
The solid curve presents the precise results, the dashed one is from
\cite{wrong}.}
\label{3}
\vspace{-10pt}
\end{figure}

We argue that the correct description of asymmetry evolution, its final
value and sign, and its effect on oscillations and on CN, is possible in  
 selfconsistent study of asymmetry, neutrino evolution and
nucleosynthesis during asymmetry's {\it whole} evolution.  
 As an illustration in fig.~3 we 
present (dashed curve) 
the calculations $\delta Y_p(\delta m^2)$ of ref.~\cite{wrong}, and 
the precisely provided calculations (solid curve).
The calculations \cite{wrong} underestimate oscillation's and asymmetry's
effects on $Y_p$ by several orders of magnitude concerning $\delta m^2$.

\section{Conclusions}
 
Precise  kinetic approach is obligatory for the study of
neutrino oscillations  in the early
Universe,  because neutrino depletion, spectrum distortion and 
oscillations generated asymmetry effects may be considerable. 
Accounting  for all oscillations effects we have reanalyzed CN
with $\nu_e \leftrightarrow \nu_s$ and updated CN
constraints on oscillation parameters.
LOW sterile solution to the solar neutrino problem is almost
completely excluded. This result is consistent with the 
last analysis of the
global data from SuperKamiokande, GALLEX+GNO, SAGE and Chlorine
experiments, which does not favour $\nu_e \leftrightarrow \nu_s$ 
LOW solution~\cite{suzuki,garcia,krastev}. 

Assumption of $Y_p=0.25$ cannot
remove CN bound on LOW solution, while small relic lepton asymmetry
$L>10^{-5}$ can evade this bound.\\

We are glad to thank P. Krastev, P. Langacker,
A. Masiero and Y. Suzuki 
for useful discussions during the Workshop. 
D.K. thanks the Organizing Committee of NOW2000 for   
the financial support of her participation.

\end{document}